# Additively manufactured polyethylene terephthalate scaffolds for Scapholunate Interosseous Ligament Reconstruction

M. Natividad Gomez-Cerezo[1], Nataliya Perevoshchikova[2], Rui Ruan[3], Kevin M. Moerman[4], Randy Bindra[2, 5], David G. Lloyd[2], Ming Hao Zheng[3,6,7], David J. Saxby[2], Cedryck Vaquette [1*].

[1] School of Dentistry, Centre for Orofacial Regeneration, Reconstruction and Rehabilitation (COR3), The University of Queensland, Herston, QLD, Australia
[2] Griffith Centre of Biomedical and Rehabilitation Engineering (GCORE), Griffith University, Gold Coast, QLD 4222, Australia
[3] Centre for Orthopaedic Research, The UWA Medical School, The University of Western Australia, Crawley, WA, 6009, Australia
[4] Biomechanics Research Centre, National University of Ireland Galway, Galway, Ireland
[5] School of Medicine, Griffith University, Gold Coast, QLD 4215, Australia
[6] Perron Institute for Neurological and Translational Science, Perth, Western Australia, 6009, Australia
[7] Australian Research Council Centre for Personalised Therapeutics Technologies, Australia

Corresponding author:

Dr Cedryck Vaquette

c.vaquette@uq.edu.au

## Abstract

The regeneration of the ruptured scapholunate interosseous ligament (SLIL) represents a clinical challenge. Here, we propose the use of a Bone-Ligament-Bone (BLB) 3D-printed polyethylene terephthalate (PET) scaffold for achieving mechanical stabilisation of the scaphoid and lunate following SLIL rupture. The BLB scaffold featured two bone compartments bridged by aligned fibres (ligament compartment) mimicking the architecture of the native tissue. The scaffold presented tensile stiffness in the range of 260 ±38 N/mm and ultimate load of 113 ± 13 N, which would support physiological loading. A finite element analysis (FEA), using inverse finite element analysis (iFEA) for material property identification, showed an adequate fit between simulation and experimental data. The scaffold was then biofunctionalized using two different methods: injected with a Gelatin Methacryloyl solution containing human mesenchymal stem cell spheroids (hMSC) or seeded with tendon-derived stem cells (TDSC) and placed in a bioreactor to undergo cyclic deformation. The first approach demonstrated high cell viability, as cells migrated out of the spheroid and colonised the interstitial space of the scaffold. These cells adopted an elongated morphology suggesting the internal architecture of the scaffold exerted topographical guidance. The second method demonstrated the high resilience of the scaffold to cyclic deformation and the secretion of a fibroblastic related protein was enhanced by the mechanical stimulation. This process promoted the expression of relevant proteins, such as Tenomodulin (TNMD), indicating mechanical stimulation may enhance cell differentiation and be useful prior to surgical implantation. In conclusion, the PET scaffold presented several promising characteristics for the immediate mechanical stabilisation of disassociated scaphoid and lunate and, in the longer-term, the regeneration of the ruptured SLIL.

**Graphical abstract**

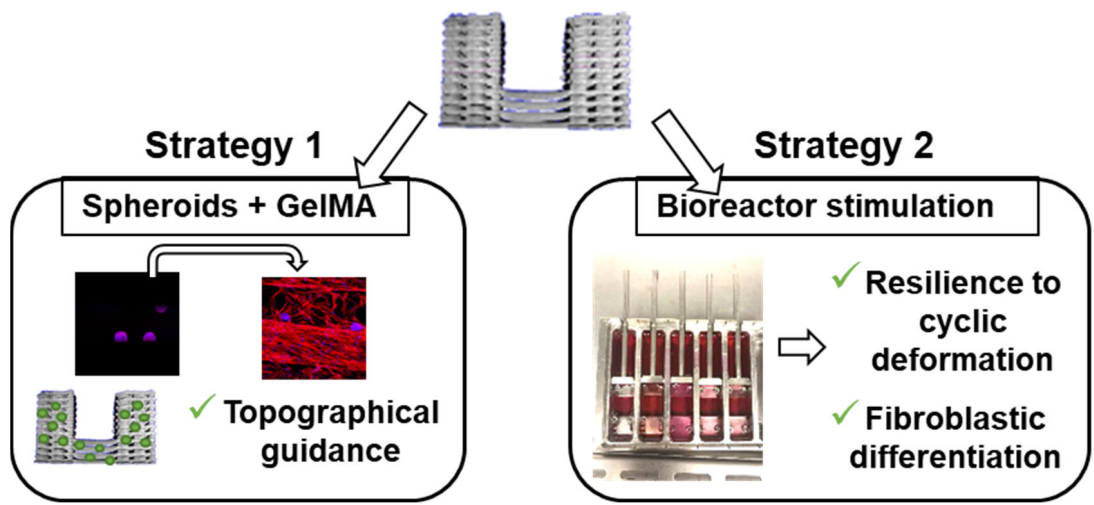

# 1. Introduction

The scapholunate interosseous ligament (SLIL) is a C-shaped ligament located in the wrist and spans the scaphoid and lunate bones. The SLIL is composed of three anatomical regions: dorsal, proximal, and volar [1]. The dorsal region is strongest and thickest, and the primary stabiliser of the scapholunate joint [2]. The SLIL, and particularly dorsal region, is the most frequently injured ligament of the wrist in young and active people. Rupture of the SLIL compromises normal wrist function, imposes significant health burden, and decreases quality of life [2-6]. Currently, effective surgical techniques for the reconstruction of partially or completely ruptured SLIL [7-11] are lacking. Since no clinical approaches results in satisfactory clinical outcomes, there is no clear consensus on the best surgical procedure. Usually, injured SLIL are treated by inserting a tendon between the scaphoid and lunate bones in order to restore their biomechanical and kinematic function. However, long-term outcomes are inadequate [12] and anatomical and biomechanical discrepancies between the graft tendon and the native SLIL results often in increased laxity of the joint over time [13]. Ultimately, when the biomechanical function of the joint cannot be restored, bone fusion [14] can be performed, resulting in complete immobilisation of the hand significantly impacting hand motion and consequently quality of life. While new surgical approaches are being developed [15], their long-term efficacy is not proven and tissue engineering strategies could be used to prepare a mechanically competent construct for surgical implantation to repair and regenerate the damaged tissue [16-18].

Recently, additive manufacture has been proposed for the engineering of bone, ligament and tendon scaffolds using conventional thermoplastic printing and bioprinting [19, 20]. Our group has also proposed a polycaprolactone (PCL) multiphasic bone-ligament-bone scaffold [17, 21] for the reconstruction the dorsal region of the SLIL. This multiphasic scaffold consisted in three compartments designed with specific architectures to mimic both the mechanical properties and hierarchical nature of the SLIL. In this bone-ligament-bone scaffold, the ligament portion (ligament-compartment) is composed of parallel fibres terminating onto plugs (bone plugs or compartments) made of orthogonally arranged matrix of fibres that interface with the scaphoid and lunate bones. *In-vivo* testing demonstrated promising performances, as evidenced by vascularisation, as well as collagen deposition in the ligament portion of the scaffold and bone formation within the bone plugs in both rodent and lapine models [17, 21]. Finite element analysis (FEA) of the 3D printed PCL scaffold showed the construct could withstand physiological movement [22]. Although these previous reports presented promising results, the short-term performance of the 3D printed biodegradable PCL scaffold has only been evaluated. In the context of a biodegrading material, the mechanical properties of the scaffold will change over time [23, 24] potentially compromising its role stabilizing scaphoid and lunate. This will be particularly the case for fast degrading polymers such as poly-glycolic acid [25] and its copolymers [26] and for aliphatic polyesters undergoing bulk degradation via autocatalytic degradation [27] resulting a rapid and drastic decrease in the mechanical properties potentially comprising the joint stabilisation. Hence, a strategy involving a non-degradable biomaterial may provide longer-lasting mechanical support and might be beneficial for the regenerative outcome.

Polyethylene terephthalate (PET) has attracted significant attention for fabrication of artificial ligaments due to its excellent mechanical and physical properties. One of the most advanced applications of PET as a biomedical device is in the replacement of anterior cruciate ligament, mainly using the ligament advanced reinforcement system LARS® [28, 29]. More closely related to the field of hand surgery, a previous cadaveric study used a modified version of the LARS ligament to replace the SLIL [30] and demonstrated mechanical stability and maintenance of physiological motion. The increased stiffness of PET compared to PCL may provide an interesting mechanical advantage to restore function of load bearing tissues, as indicated by use when replacing the anterior cruciate ligament. However, PET is bio-inert in nature and surface modification is generally required for enhancing biological interactions [29, 31, 32] in order to achieve proper tissue integration and regeneration.

The objective of the present work was to evaluate the performance of a PET 3D printed bone-ligament-bone scaffold for the reconstruction of the SLIL and assess its suitability for achieving scapholunate joint stabilisation. This study reports on the printing optimisation of the PET bone-ligament-bone scaffold where the ligament portion of the scaffold was assessed using mechanical testing and inverse finite element analysis (iFEA) to describe the PET scaffold's isotropic hyperelastic properties and scaffold-ligament fibre stresses. In addition, a proof of concept for increasing the biofunctionality of the scaffold is demonstrated using two different strategies with (i) cell spheroids delivered in a hydrogel, or (ii) direct cell seeding and maturation in a bioreactor to enhance extracellular matrix formation prior to implantation.

## 2. Materials and methods

## 2.1. Scaffold manufacturing

First, the manufacturing process was optimised to select the most appropriate printing conditions. Various printing parameters were investigated (Table 1) using a 3D-Flashforge Inventor 3D Printer, Dual Extruder machine (Zhejiang Flashforge 3D Technology Co. Ltd. Jinhua City, Zhejiang Province, China) with PET filaments (AURARUM, Australia).

The Bone-Ligament Bone scaffold consisted in two bone compartments bridged by a ligament compartment. Similar to our previous study, a U-shape bone-ligament-bone scaffold was manufactured [17, 21]. A STL file featuring the shape of the scaffold was created using the open-source Autodesk Tinkercad software, which was then sliced at a 360 μm layer height using Simplify3D®. The scaffold design was assigned a 40% infill and 0.7 mm fibre spacing, with 4 layers consisting of 7 fibres per layer, totalling 28 fibres in the ligament-compartment. The layer-to-layer orientation was 0/0° (i.e., parallel) for the ligament compartment. The bone compartments were assigned a 50% infill, 0.7 mm fibre spacing and 0/90° (i.e., orthogonal) layer-to-layer orientation. The resulting file was transferred to the 3D Flashforge Inventor 3D Printer, Dual Extruder, which printed scaffolds using a 400 μm diameter nozzle with different temperatures and velocities (Table 1). The effect of these printing parameters on mechanical properties of the prints were evaluated using various experimental tests.

**Table 1.** Summary of 3D printer processing parameters.

| Sample name | Temperature ($^o$C) | Speed (mm/s) |
|---|---|---|
| 200-300 | 200 | 300 |
| 200-450 | 200 | 450 |
| 200-600 | 200 | 600 |
| 205-300 | 205 | 300 |
| 205-450 | 205 | 450 |
| 205-600 | 205 | 600 |

Once the scaffolds were manufactured, their structures were examined using scanning electron microscopy and microcomputed tomography. The scaffolds also underwent mechanical assessment using experimental mechanical testing and finite element analyses to identify their material and mechanical properties.

## 2.2. Scanning electron microscopy

Scanning electron microscopy (SEM) was carried out using a JSM F-7001 microscope (JEOL Ltd., Tokyo, Japan), operating at 5 kV and 40 mm of working distance. The scaffolds were mounted onto SEM stubs and carbon coated in a vacuum using a sputter coater (Balzers SCD 004, Wiesbaden-Nordenstadt, Germany).

## 2.3. Scaffold porosity by microcomputed tomography

Microcomputed tomography (MicroCT) imaging was performed using a Skyscan 1272 (Bruker, Billerica, MA, USA) to determine the porosity of the U-shaped PET scaffolds (n=3). The scanning parameters were: 50 kV X ray voltage, 200 µA current, 150 ms exposure time, 10 µm isotropic voxel size, 0.3° rotation step, 2 frame averaging, 4 × 4 binning without filter. The datasets were reconstructed with NRecon Version 1.7.3.1 (Bruker, Billerica, MA, USA) and InstaRecon Version 2.0.4.2 (University of Illinois, Champaign, IL, USA) using a cone beam reconstruction (Feldkamp) algorithm with the following corrections applied: ring artefact reduction, smoothing, beam hardening, and post-alignment. The collected data from the CT scans were processed and analysed using the CTan software version 1.19.11.1 (Bruker, Billerica, MA, USA).

## 2.4. Scaffold mechanical testing

For the purpose of mechanical testing, a modified version of the different scaffolds was created. In this, the upper portions of the scaffold's bone plugs were removed to create a rectangular construct with the ligament compartment in the middle with 28 fibres (7 fibres/layer, total 4 layers) and the bone compartments at the ends. The final rectangular PET 3D printed scaffolds were 15 mm long, 5 mm wide, and 3 mm high, and were assessed via both uniaxial tensile and three-point bend tests.

Uniaxial tensile testing of scaffolds was performed using a 5543 Instron Microtester fitted with a 500 N load cell. The scaffolds were tested in PBS bath at 37 ºC using a crosshead speed of 20 mm/min and a 5 mm gauge length. The samples (n=6) were placed into the grips, ensuring the ligament-scaffold fibres were parallel to the longitudinal axis between the grips and pre-loaded with 0.1 N.

Prior to mechanical testing, the bone matrix ends of the scaffolds were filled with a fast-setting epoxy resin (Selleys Araldite 5 min, Australia) to stiffen the bone compartments before being subsequently inserted in the grips. This prevented compression and stress concentration at the border between the bone and ligament-scaffold compartments.

The force (N) and displacement (mm) were recorded during the tests. The averaged force and displacement from number of runs on each sample are reported. The engineering stress (MPa) was calculated based on the sum of area of the fibres in the ligament-scaffold part. Engineering strain was calculated based on the elongation of the ligament part during the tests. The stiffness (N/mm) was calculated from the linear part of the elastic region of the force-displacement curves. The ultimate force (N) was determined from the force-displacement curves in order to directly compare with the SLIL biomechanical properties reported in the literature. The tensile Young's modulus (MPa) was calculated from the linear portion of the stress-strain curve. The ultimate tensile strength was derived from the stress-strain curve at the first point of failure (that is the first abrupt decrease in stress or load). Thereafter, a scaffold was selected based on morphological characterisation and mechanical properties and used for the rest of the study.

## 2.5. Finite element analysis

An inverse finite element analysis (iFEA) was used to identify PET material properties followed by assessment of applied stresses along the long axis in a single fibre ligament-scaffold. The best printing conditions were used to print the single fibre ligament-scaffold for mechanical testing. All data processing and visualization were performed using custom MATLAB R2021a (The Mathworks Inc., Natick, MA, USA) code and the open-source MATLAB toolbox GIBBON ([33] https://www.gibboncode.org). Following material property fitting, an FEA was implemented on the entire ligament-scaffold model using the open source software FEBio Version 3 (Musculoskeletal Research Laboratories, The University of Utah, USA, http://febio.org, [34].

The first step in the FEA was meshing the ligament-scaffold model which was accomplished using hexahedral elements, with 9504 elements in each single fibre. The mesh elements had homogeneous isotropic and hyperelastic constitutive properties. Therefore, the hyperelastic behaviour was defined by the following coupled Neo-Hookean hyperelastic strain energy density formulation, $\Psi$:

$$\Psi = \frac{\mu}{2}(I_1 - 3) - \mu \ln J + \frac{\lambda}{2}(\ln J)^2 \qquad (Eq.1)$$

where, $I_1$ is the first invariant of the right Cauchy-Green tensor and $J$ is the determinant of the deformation gradient tensor. The parameters $\mu$ and $\lambda$ are the Lamé parameters, which can be related to the Young's modulus (Young's modulus), and $v$ (Poisson's ratio) as follows.

$$\mu = \frac{E}{2(1+v)} \tag{Eq.2}$$

$$\lambda = \frac{vE}{(1+v)(1-2v)} \tag{Eq.3}$$

### 2.5.1. Fitting constitutive equations parameters

An iFEA technique was used to identify the PET scaffold's material parameters in the constitutive equations (Eq. 1). In this, the PET material's Young's moduli ($E$) were found by optimization, while Poisson's ratio was set constant at $v=0.4999$ to approximate the realistic incompressible behaviour of PET. The moduli, $E$, were adjusted until the simulated force-displacement responses best matched (i.e., minimal error) those measured during uniaxial tensile testing experiments. This inverse parameter identification used FEBio to solve the FEA simulations which were nested within Levenberg-Marquardt based optimization, all coded in MATLAB [35].

Parameter optimization followed the procedure described next. Using either initial or best updated material properties, FEBio solved for the force applied to a single-fibre FEA model using displacement inputs that simulated experimental values from the uniaxial tensile testing. The estimated single-fibre applied force was then multiplied by 28 to produce the total simulated force applied to the 28-fibre ligament-scaffold ($F_i^{sim}$). This assumed the displacement applied to one end of the ligament-scaffold was the same across for all fibres, and, therefore, the same force equally distributed across all fibres. The simulated force ($F_i^{sim}$) was then compared to experimental forces ($F_i^{exp}$) from the uniaxial tensile testing, which was formulated as the optimization objective function $\phi(E)$ as the summed squared errors between $F_i^{sim}$ and $F_i^{exp}$, i.e.,

$$\phi(E) = \frac{1}{n}\sum_{i=1}^{n}(F_i^{exp} - F_i^{sim})^2 \tag{Eq. 4}$$

Where, $n$, is the number of discrete experimental data points from the uniaxial tensile test. The optimization was deemed to have converged if either $E$ or $\phi(E)$ did not vary by more than 0.01.

After finding the optimal Young's modulus, the simulated 3D stresses distribution in a single fibre ligament-scaffold was visualized.

**2.6. Enhancement of polyethylene terephthalate scaffold biofunctionality**

As a proof of concept for enhancing the bioactivity of the scaffold, two treatments were trailed. These used PET scaffolds treated with radio-frequency oxygen plasma, and (i) injected with a Gelatine Methacryloyl solution containing human mesenchymal stem cell spheroids in order to deliver cell to aid regeneration as conventionally performed in cell-based therapies, or (ii) tendon-derived stem cells directly seeded in the scaffold and application of cyclic biomechanical stimulation for ECM maturation.

**2.7. Radio-frequency oxygen plasma treatment**

Radio-frequency oxygen plasma treatment was carried out to increase the hydrophilicity of the rectangular PET scaffold printed with optimized printer parameters and improve its interaction with biological entities. This facilitated the homogeneous infiltration of the cellularised viscous GelMA solution and fluid circulation in the bioreactor approach. Plasma treatment was performed using a modified protocol previously reported for polycaprolactone 3D-printed constructs [36]. The plasma was created with an inductively coupled radio frequency (RF) generator, operating at a frequency of 27.12 MHz and an output power of about 30 W. The samples were exposed to an oxygen flow of 10.6 mL/min and an argon flow of 4.1 mL/min for 4 minutes. The prepared samples were mechanically tested and compared to untreated rectangular PET scaffolds printed with the same optimal printing parameters.

X-ray photoelectron spectrometry was used to evaluate the surface of the rectangular scaffolds before and after plasma treatment, using a Kratos Axis ULTRA x-ray photoelectron spectrometer incorporating a 165 mm hemispherical electron energy analyzer. The conditions and analysis were similar to the previously reported [37], using CasaXPS version 2.3.14 software for peak fitting of the high-resolution scans.

**2.8. Human bone marrow mesenchymal stem cells**

Human bone marrow mesenchymal cell (hMSC) spheroids were encapsulated in Gelatine Methacryloyl and injected into the scaffolds. Three different amounts of spheroids (300, 600 and 1200) were injected in order to find the best amount of spheroid seeding.

In the cell culture preparation, hMSC (PT-2501, Lonza Australia, Mt Waverley, VIC) were expanded in Dulbecco's modified Eagle's medium (DMEM, Gibco) until passage 5, changing the media every 2-3 days. Basal cell culture media was a combination of Dulbecco's modified Eagle's medium (DMEM, Gibco) with 10% fetal bovine serum and 1% Penicillin Streptomycin (Gibco).

### 2.8.1. Human bone marrow mesenchymal cell spheroid preparation

hMSC spheroids were generated using the AggreWell system following the manufacturer's instructions. Two mL of Rinsing Solution (Stemcell Technologies) was pipetted into each well of the Aggrewell$^{TM}$400Ex plate (Stemcell Technologies) and centrifuged at 2000 g for five minutes. The Rinsing Solution was removed, and the wells were washed with complete medium and 0.5 mL of complete media was added to the well. hMSC (Lonza Australia Pty Ltd) were suspended in basal media at a concentration of 500 cells/µL to form 500 cells/ spheroid. The cell suspension was placed in the well and pipetted gently to ensure homogeneous dispersion. The AggreWell Plate was incubated at 37 °C, 5% $CO_2$ for 24 hours to induce spheroid formation.

### 2.8.2. Spheroid encapsulation in Gelatin Methacryloyl crosslinking

Porcine, lyophilized Gelatin Methacryloyl (GelMA) (Gelomics, Brisbane, Australia) was dissolved in DMEM at 37 °C to obtain a 10% wt/vol solution, and lithium phenyl-2,4,6-trimethylbenzoylphosphinate (LAP) was added as a photo-initiator at a 0.05% (wt/vol) concentration. Different amounts of spheroids (300, 600 and 1200) were suspended in 60 µL of GelMA and injected into each bone and ligament compartments of the scaffold. Thereafter, the scaffolds were briefly maintained at 4 °C for 2 min to allow for rapid thermal gelation of the GelMA preventing spheroid sedimentation prior to non-reversible photo-crosslinking using a custom made 15 x 12 $cm^2$ light emitting diode panel with a total power of 20 W and 405 nm wavelength at 1 cm curing offset for 2 min. Finally, the scaffolds were placed in 2 mL medium and cultured over 10 or 21 days.

### 2.8.3. Cell viability

The spheroid cells viability was assessed at 24 hr and 10 days post-formation. To study cell viability after encapsulation in the GelMA and injection into the PET scaffold, a Live/Dead® assay was performed by staining the cells of the spheroids with fluorescein diacetate (FDA- green channel for living cells) or propidium iodide (PI- red channel for dead cells) after 1, 4, 7 and 10 days (n = 2 for each timepoint). The free spheroids and those embedded in the scaffolds were washed twice in PBS, then incubated with FDA

(0.8 U/ml) and PI (5 μg/ml) in PBS for 10 min at 37 °C under 5% $CO_2$. The scaffolds were then rinsed twice in PBS and imaged immediately after. The cells were visualised under an Eclipse Ti Confocal Microscope (Nikon, USA) at excitation/emission wavelength 488/530 nm for FDA and 561/620 nm for PI.

### 2.8.4. Cell proliferation

Metabolic activity of the spheroids in the scaffold at three seeding densities (300, 600, and 1200 spheroid per scaffold) was studied at 1, 3, 7, and 10 days using Alamarblue® (ThermoFischer) assay. In brief, scaffolds (n=3 for each timepoint) were placed in new wells, then immersed in 100 μL of the resazurin solution with concentration (10% vol/vol) for 4 hr at 37 °C under $CO_2$ (5%) atmosphere. Then, the Alamar Blue solution fluorescence was measured by a Tecan infinite M200 Pro spectrophotometer in duplicates. The excitation wavelength used was 560 nm and emission wavelength used was 590 nm. In addition, the metabolic activity of scaffolds seeded with 1200 was studied up to 21 days. After Alamar Blue exposure, the scaffolds were rinsed in medium and further cultured until the next time point.

### 2.8.5. Morphological studies by confocal laser scanning microscopy

Confocal microscopy was performed to visualise the morphology of the GeMA-encapsulated spheroids injected into the scaffolds at 1, 4, 7, and 14 and 21 days (n=2). Each scaffold was rinsed twice in phosphate buffer solution (PBS) and fixed in 4% (w/v) paraformaldehyde in PBS at 37 °C for 4 hr. Then, the samples were incubated for 5 min with Triton 0.1% (w/v) at room temperature to permeabilise the cells. After, the scaffolds were rinsed with PBS and incubated with 400 μL, 6-diamidino-2-phenylindole (DAPI) (DAPI, Vector Laboratories, Burlingame, CA, USA) (1:1000) and Atto 565-conjugated phalloidin at a concentration of 0.165 μM (Molecular Probes) for 30 min and subsequently imaged using a Nikon Confocal Microscope (Eclipse- Ti, U.S.A).

### 2.8.6. Bioreactor with tendon-derived stem cells to increase scaffold biofunctionality

A bioreactor applying cyclic mechanical deformation was utilised to assess the mechanical performance of the scaffold under fatigue and to investigate the effect of mechanical stimulation on the cells directly seeded on the PET scaffold. The bioreactor was designed to stimulate ligament differentiation of tendon-derived stem cells (TDSC) by applying a loading regime of 6% strain at 0.25 Hz for 8 h, followed by 16 h rest for 6 days according to our previously optimised protocol [38]. TDSC were utilised in order to maintain consistency with our previous publications using this optimised bioreactor approach [38, 39].

In brief, a uniaxial stretching system was used to stimulate the TDSC differentiation into ligament-like tissues [38, 39]. The followed protocol to isolate and seed TDSC cells has been previously described [39]. In brief, mice TDSC cells were isolated from 6–8-week-old C57BL/6 mice and isolated in complete media (α modified Eagle's medium, supplemented with 10% foetal bovine serum, 100 U/ml penicillin, 100 μg/ml streptomycin) in T-25 flask at 37 °C in a humidified atmosphere, containing 5% $CO_2$ [38]. After 7–10 days of culture, TDSC were trypsinized and transferred into T75 flasks at density of 400 cells/cm$^2$ and the medium was changed every 3 days. Then, TDSC at passage 4 were seeded with density of $1.5 \times 10^6$ cells per scaffold (resulting in a cell seeding density of $6.6 \times 10^6$ cells/cm$^3$). While the dimension of the native ligament is around 5mm, the scaffolds were printed with dimensions 15 x 5 x 3 mm$^3$ in order to adjust to the dimensions of the bioreactor and to ensure the production of detectable level of proteins. The cell seeding was performed by pipetting 1000 μL of medium containing $1.5 \times 10^6$ cells onto the scaffold. The scaffolds were cultured for 3 days after which they were mounted into the bioreactor. Subsequently the bioreactor was filled with complete medium (α modified Eagle's medium, supplemented with 10% fetal bovine serum, 100 U/ml penicillin, 100 μg/ml streptomycin), and then underwent the aforementioned loading regime.

Proteins were extracted from the PET scaffolds after 6 days in the bioreactor, as well as from control PET scaffolds (n=6). Control scaffolds were seeded with cells but not cultured under mechanical stimulation. All samples were immersed in a Radioimmunoprecipitation buffer (Tris-HCl 50 mM PH=8, NaCl 0.75 M, NP-40 0.5%, EDTA 5 Mm, 0.1 Mm PMSF, a 1 x CPI made from 25X and a DNAse I 500 μG/ml was added to the buffer) for 48 hr at 4 °C. The lysates were centrifuged 12,000 rpm for 10 min, the supernatant was collected and quantified using a Bradford assay. We performed a Sodium Dodecyl Sulfate-Polyacrylamide Gel Electrophoresis (SDS-PAGE) with 10 μg of proteins for all samples. An SDS-PAGE gel comparing Tenomodulin (TNMD) (Anti-tenomodulin antibody ab203676 from Abcam, UK) proteins and Collagen 1 (COL1) (Anti-Collagen I antibody ab90395 from Abcam, UK) indicated the presence of bands that match their molecular weight. The protein quantification was then analysed by ImageJ.

**2.9. Statistical analysis**

The statistical difference was analysed by one-way ANOVA followed by a Tukey post-hoc analysis using GraphPad Prism 8 software. A probability level p<0.05 was considered as significant with *, ** and *** representing p<0.05, p<0.01 and p<0.001, respectively. When only two groups were compared a t-test was utilised in order to determine statistical significance (p<0.05).

# 3. Results

## 3.1. Effect of 3D-printing conditions on scaffold morphology

The structure of the PET scaffolds varied with different printing temperatures and speeds (Figure 1). An increase of 5 °C reduced both the alignment of the filaments within the ligament compartment and the printing quality. Cross-sectional images showed PET fibres underwent notably more sagging in the ligament compartment when printing temperature was higher. Overall, a small increase in printing temperature negatively affected scaffold structure.

Printing speed played an essential role in the filament alignment. Compared to scaffold printed at lower temperatures, scaffolds printed at 300 mm/s displayed poorer fibre alignment (as seen in both top view and cross section), impairing overall scaffold structure. This effect was enhanced by increased printing temperature, whereby disruption in filament alignment resulted in uncontrolled fibre placement. Interestingly, as printing speed was increased the quality of the printing and its reproducibility also improved. The morphological analysis showed the best printing conditions were 200 °C and 600 mm/s, allowing a high control over printing accuracy and overall scaffold structure.

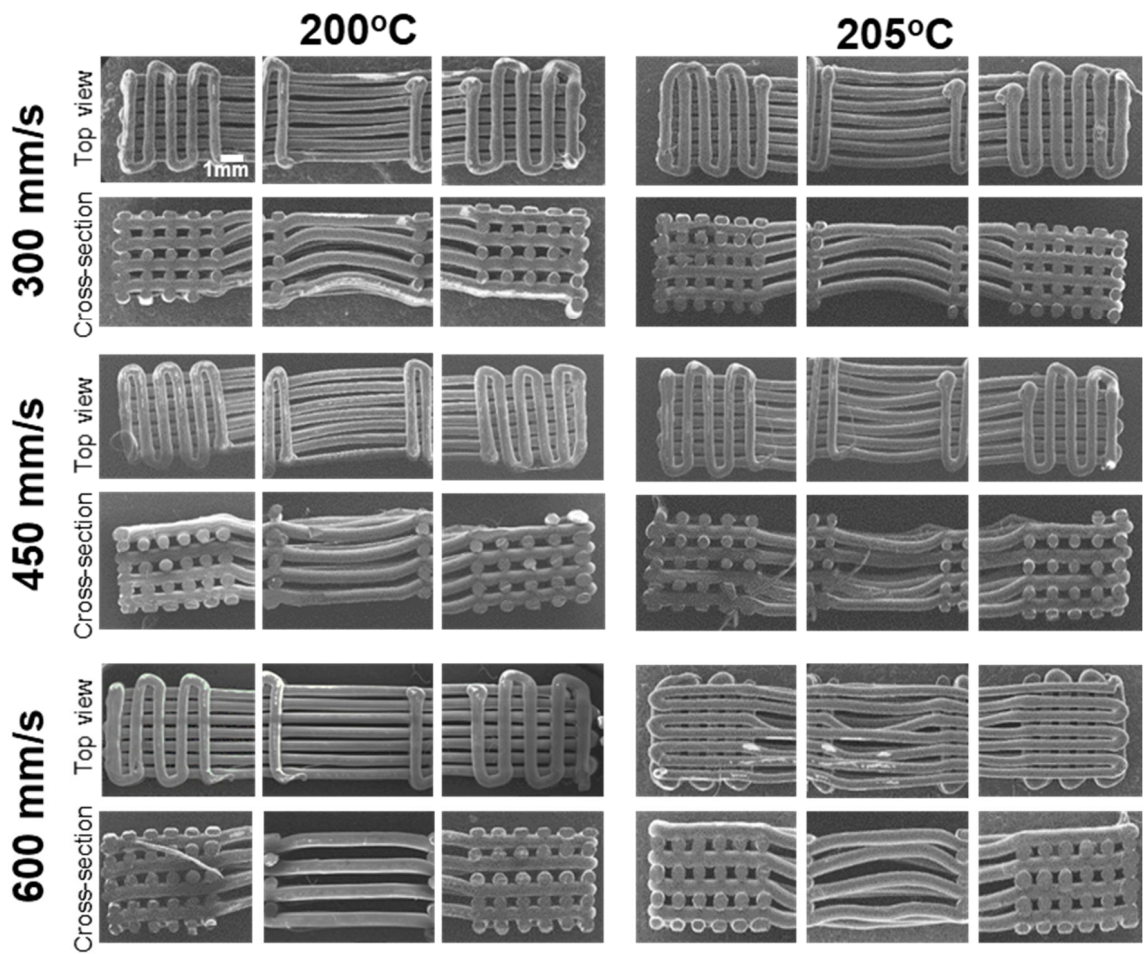

**Figure 1.** Scanning electron microscopy figures of rectangular 3D-printed scaffolds at various conditions.

### 3.2. Effect of 3D-printing conditions on mechanical properties

The tensile mechanical properties of the construct were assessed under quasi-static conditions. The 3D printed scaffold showed an engineering stress–strain curve typical of polymeric material: an initial linear region followed by a yield point (Figure 2A and D) and a plastic deformation region (not shown). Stress-strain curves showed similar shapes for all samples, however, several differences were observed in various mechanical parameters. Indeed, the lowest printing temperature seems to produce specimens with higher stiffness and young modulus, although it only reached statistical significance between the 200- 450 and the 205-300 groups. Similarly, there was a slight effect of the printing speed on the mechanical properties as lower speed resulted in lower stiffness and young modulus although it only reached statistical significance between the 200-300 and the 200-450 groups. There was not marked difference in the ultimate load and ultimate stress. This indicated that the printing temperature and speed had only minimal impact on the mechanical properties of the scaffolds. As such all conditions presented Young's, Stiffness ranging from 200-275 N/mm, Ultimate Load around 110N, Modulus ranging from 800 to 1000 MPa, and ultimate stress around 42 MPa.

These results, together with the scanning electron microscopy results confirmed that scaffolds printed at 200 °C and 600 mm/s allowed for high control over internal structure and mechanical properties. Therefore, scaffolds were printed at these conditions and used for the rest of the study.

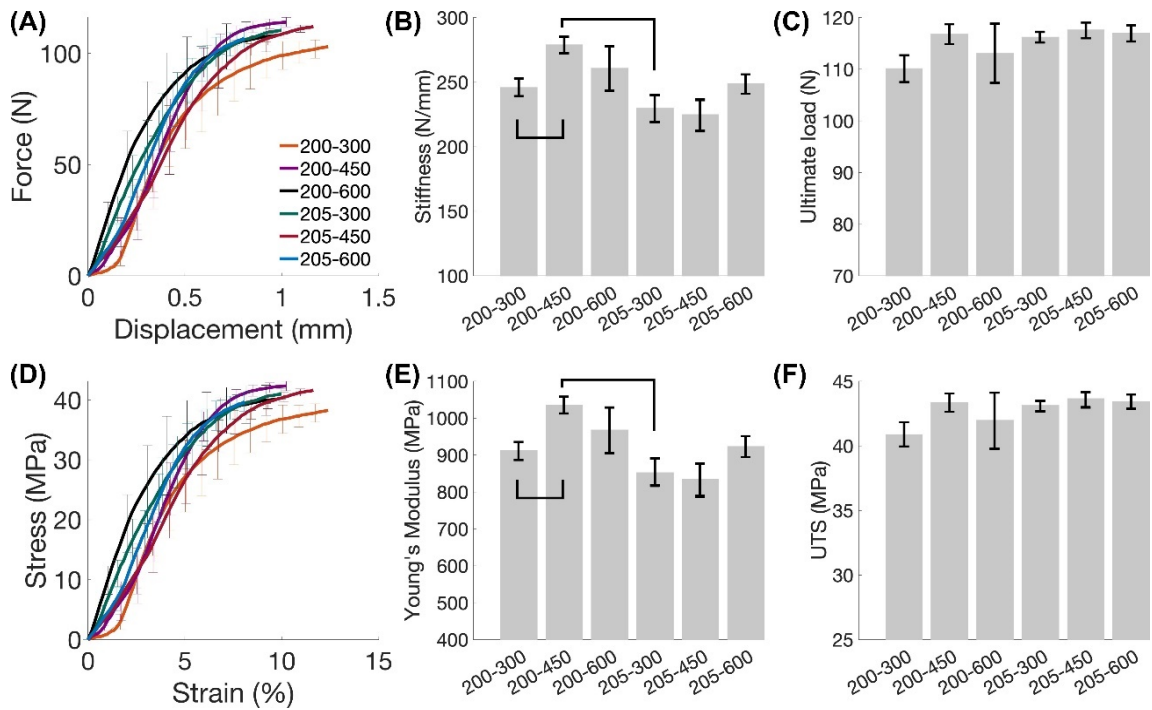

**Figure 2.** Tensile properties of the scaffolds printed at different temperatures and speeds. (A) representative elastic region of force vs displacement, (B) stiffness and (C) ultimate load, (D) elastic region of engineering stress vs strain curves for all printing conditions, (E) Young's modulus; (F) ultimate tensile strength (n=5). Bars denote statistical significance (p<0.05).

### 3.3. Finite element analysis

The FEA model of the scaffold well fit the linear elastic region of the experimental data (Figure 3A), with an optimized Youngs Modulus ($E$) of 741.23MPa and a Poisson's ratio of 0.4999. The Cauchy stress ($\sigma_{zz}$) in the loading direction distribution within a single fibre are visualised in the Figure 3B. The higher stresses concentrate in the middle and lessened towards the ends.

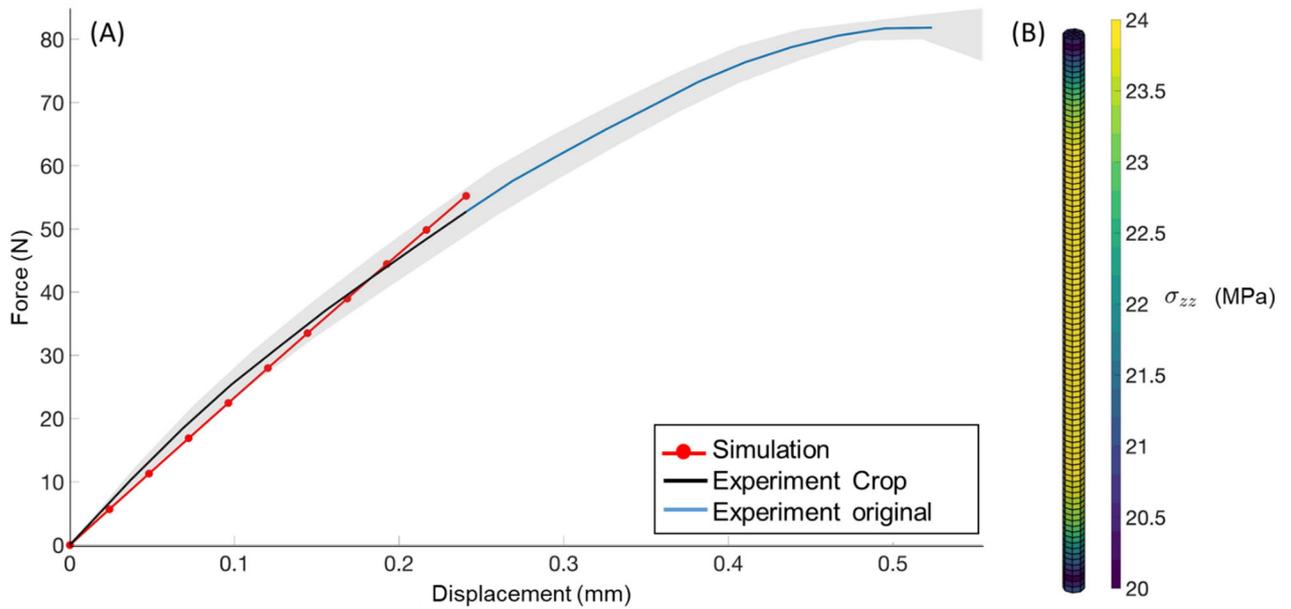

**Figure 3:** (A) Force-displacement curve and (B) calculated stress distribution (MPa) using the material constants from the optimisation of the elastic properties.

### 3.4 Morphology of the bone-ligament-bone scaffold

The SEM images confirmed a high manufacturing control over porosity and macrostructure of the scaffolds printed (bone and ligament compartments) at 200 °C and 600 mm/s (Figure 4). The microCT quantified porosity of the scaffolds printed under these conditions matched the theoretical printing porosity (Figure 4F). The bone plugs had a 50% theoretical porosity confirmed by a measured ~55%, while the ligament compartment had a 60% theoretical with a similar measured porosity of ~65% which is consistent with our previous study [17]. The pore size in the bone compartment was 250 μm by 500 μm and therefore suitable for tissue infiltration and rapid neovascularisation.

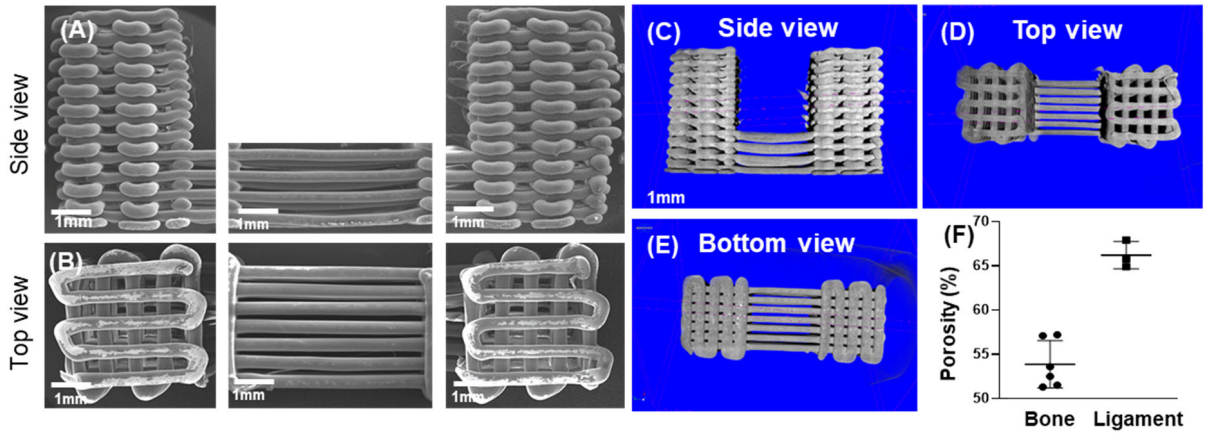

**Figure 4.** Scanning electron microscopy images of the scaffolds' structure when printed at 600 mm/s and 200°C with bone plugs included (A) side view, (B) top view, and microCT reconstruction (C) side view, (D) top view, (E) bottom view, and (F) porosity of ligament and bone compartments (n=3).

### 3.5. X-ray photoelectron spectrometry

The atomic concentration of oxygen on the PET surface was 7% higher after plasma treatment, increasing the O/C ratio from 0.23 to 0.36 (Table 4).

**Table 4**. Chemical composition by X-ray photoelectron spectrometry of PET surface before and after treatment in oxygen plasma. C - Carbon, O - Oxygen.

| Sample | C (At%) | O (At%) | O/C |
|---|---|---|---|
| PET | 81.0 | 19.0 | 0.23 |
| PET-plasma treated | 73.7 | 26.3 | 0.36 |

The X-ray photoelectron spectra of the PET and plasma treated-PET scaffolds confirmed the formation of oxygen-containing (C-O and C=O) functional groups by plasma treatment (Figure 6A and 6C). Both C 1s spectra presented three peaks, before plasma treatment C=O (288.9 eV), C–O (286.5 eV), and C–C/C–H (284.1eV) and after, C=O (288.6 eV), C–O (287.8 eV), and C–C/C–H (285.1 eV), showing signal displacement with larger peaks. In addition, the amount of C-O and C=O increased about 7% and about 3%, respectively, confirming the increase in the amount of oxygen on the surface. The spectra of untreated PET exhibited two peaks assigned to the O=C (530.5 eV) and O–C (529.1 eV) bonds in the ester group. The O 1s spectra of plasma treated PET spectra were very similar O=C (530.1 eV) and O–C (529.2 eV), each having similar amount of chemical bonds before and after the plasma treatment.

As the plasma treatment modified the surface chemistry of the scaffold, it also significantly impacted on the mechanical properties. As shown in Figure 5 E-J), there were significant decreases in the various parameters considered. Figure 5E shows a drastic difference in the force/displacement and stress/strain curves which consequently resulted in a significant decrease in the stiffness (~35% decrease for both temperatures) (Figure 5 F), ultimate load (25-30% for both temperatures) (Figure 5 G), Young's Modulus (~ 30% decrease for both temperatures) (Figure 5 H) and Ultimate Tensile Stress (25-30% for both temperatures) (Figure 5 J).

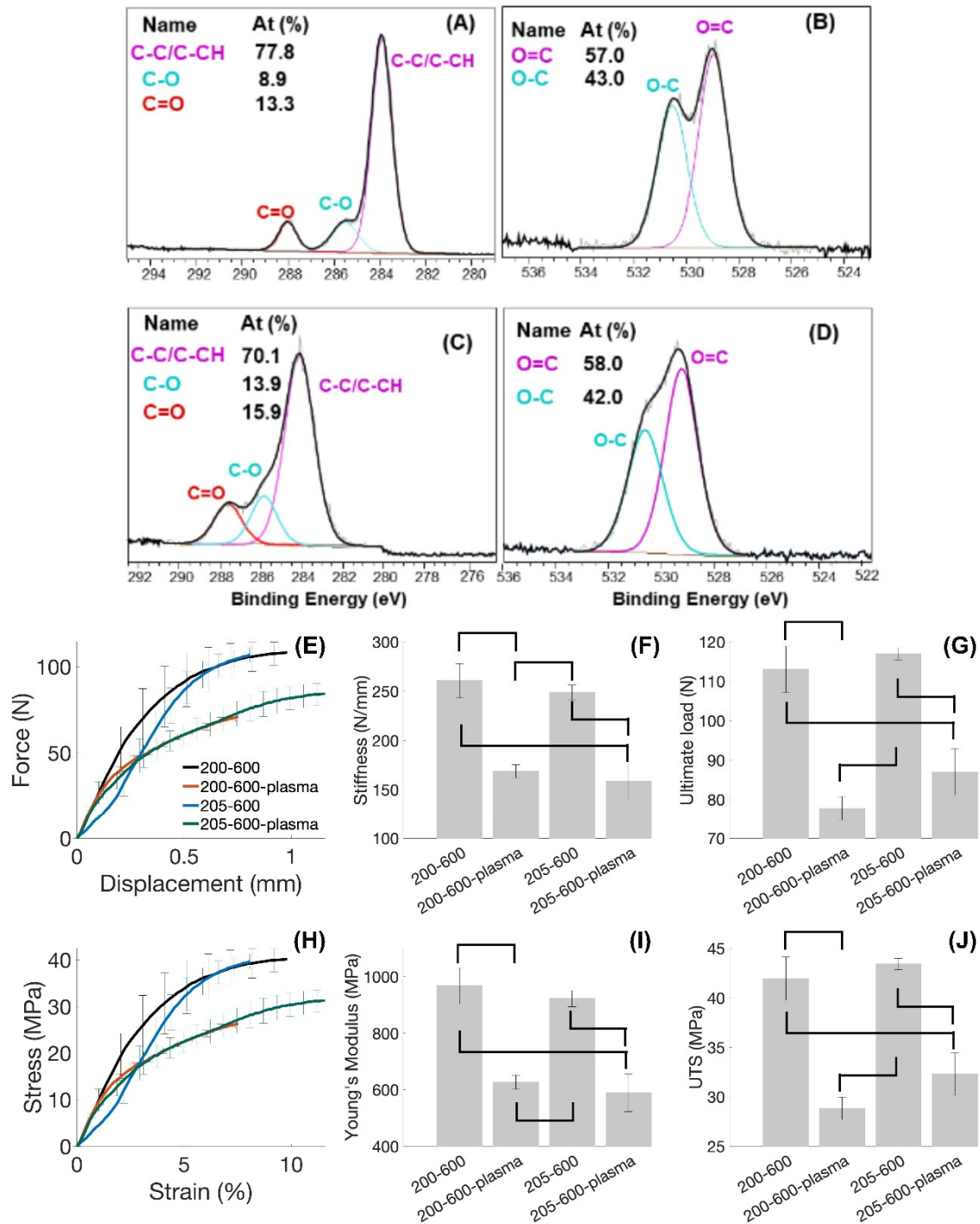

**Figure 5.** Fitting X-ray photoelectron spectroscopy spectra of (A) Carbon and (B) Oxygen of the PET scaffolds, and (C) Carbon and (D) Oxygen of plasma treated PET scaffolds. Also shown are (E) representative elastic region of force vs displacement, (F) stiffness, (G) ultimate load, (H) engineering stress vs strain curves, (I) Young's modulus, and (J) ultimate tensile strength of scaffolds at two different printing conditions before and after plasma treatment (n=5). Bars denote statistical significance ($p<0.05$).

### 3.6. *In vitro* studies

*hMSC spheroids studies*

The imaging of the Aggrewell culture plates showed the formation of hMSC spheroids 24 h after seeding with different sizes ranging from 150 to 270 μm in diameter (Figure 6A, B and C). The aim of the spheroid methodology was to create small diameter spheroids to minimise the formation of a necrotic core. After 1 day, live-dead assay revealed high hMSC cell viability, despite presence of dead hMSC cells in the core of the spheroid. Our result indicates that despite the small diameter, diffusional issues must have occurred since some cells death was observed, although this was quite limited 24hrs post formation. After 10 days in culture, spheroids had attached to the bottom of the culture plate, resulting in hMSC cell migration across this surface. The cells attached on the plate demonstrated high viability whereas the core of the spheroid displayed high cell death (Figure 6C).

When injected into the scaffolds, the spheroids encapsulated in GelMA (Figure 6D) showed similar characteristics to those on the culture plate (Figure 6D). The results revealed the outer layer of the spheroid were highly viable whereas the presence of a necrotic core was observed, similar to the free-floating spheroids (Figure 6D). As early as 4 days post-encapsulation in GelMA, cell migration into the hydrogel was observed and these cells presented a high viability similar to the observations of the spheroid made in the tissue culture plates. At longer culture timepoints, more cells colonised the hydrogel and started following the architectural features of the PET 3D-printed scaffolds and this phenomenon was observed regardless of the compartment. Consequently, these cells presented high viability whereas the centre of the remaining spheroids consisted of dead cells. The increased seeding densities of spheroids has corresponding increase in cell metabolic activity (Figure 6E) following a proportional relationship. Since the highest spheroid seeding density resulted in enhanced metabolic activity, this condition was selected to be further cultured until 21 days (Figure 6F). This demonstrated that prolonged culturing had further increased metabolic activity indicating the cells were alive and proliferating.

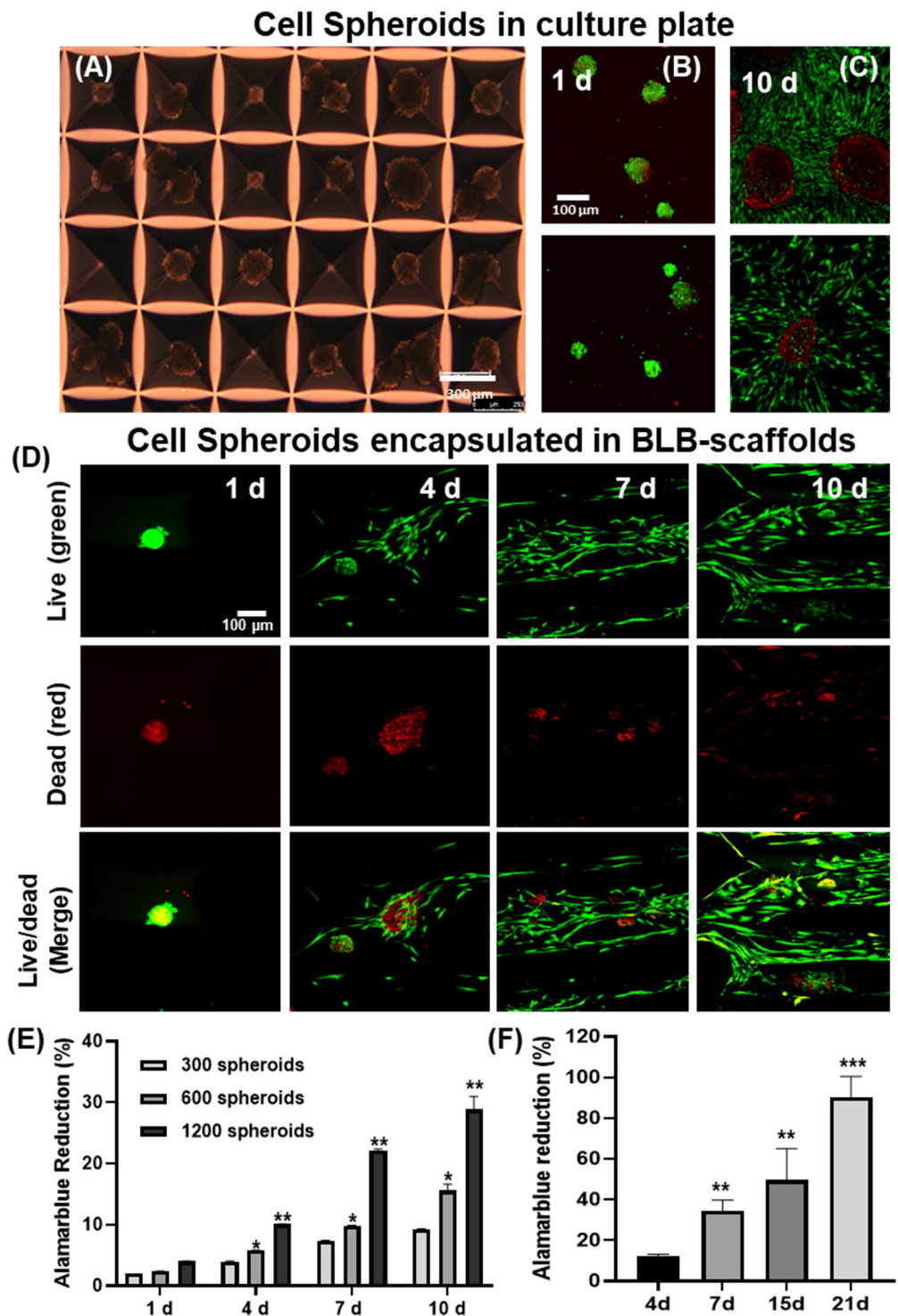

**Figure 6.** (A) hMSC spheroids formed in AggreWell400Ex culture plate, cell viability of the spheroids cultured in 24-well plate of the cells at (B) 1 day and (C) 10 days. (D) Cell viability of the spheroids in GelMA injected in the scaffolds (1200 spheroids). (E) Spheroid cell metabolic activity over 10 days of culture in the scaffolds at different initial

spheroid numbers in the scaffold (n=3). (F) Cell metabolic activity at 21 days in scaffolds injected with 1200 spheroids (n=5). Stars denote statistical significance (p<0.05).

Cellular morphology, investigated by staining the nucleus and cytoskeleton, confirmed the observations made with the live-dead assays; the cells that migrated out of the spheroid gradually colonised the hydrogel and followed the architectural features of the 3D-printed scaffold (Figure 7).

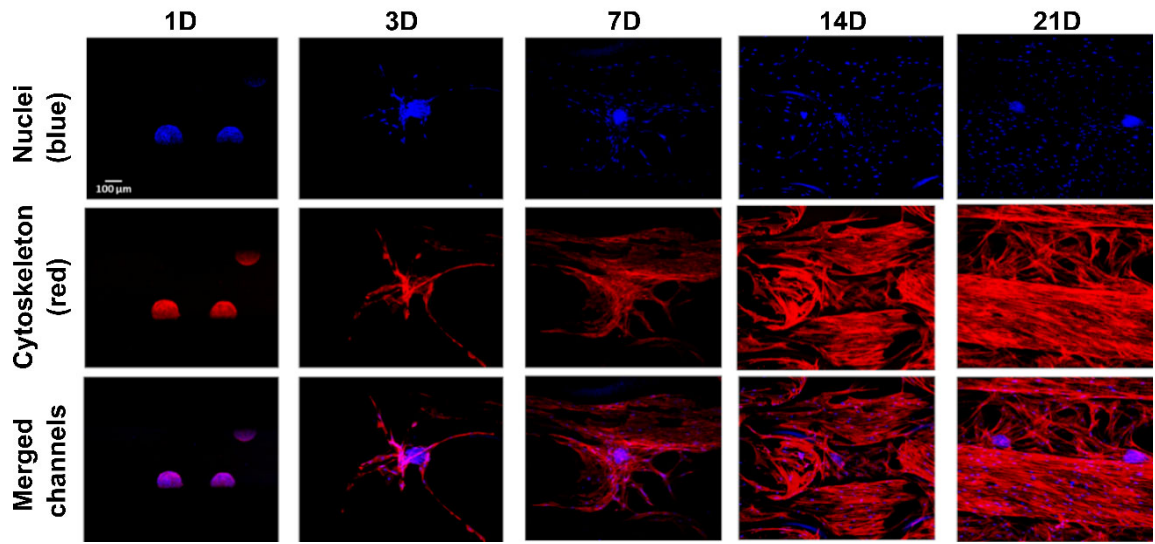

**Figure 7.** Confocal laser scanning microscopy images of F-actin/nucleus staining showing cellular morphology and spreading in the GelMA hydrogel injected in the 3D printed scaffold after 1, 4, 7, 14, and 21 days of culture (n=2).

*Bioreactor studies*

Subsequent to the mechanical stimulation received in the bioreactor, the scaffold's mechanical properties were assessed using tensile testing. There were non-significant differences in the ultimate load, stiffness, and Young's modulus after 6 days loading regime (Figures 8A-F). No permanent plastic deformation (i.e., elongation) was observed in the scaffolds (Figure 8G) after the cyclic bioreactor loading. Interestingly, the Western blot showed significantly increased relative expression of TNMD in the PET-scaffold, while the COL1 remained the similar in both groups, after 6 days in the bioreactor (Figure 8H).

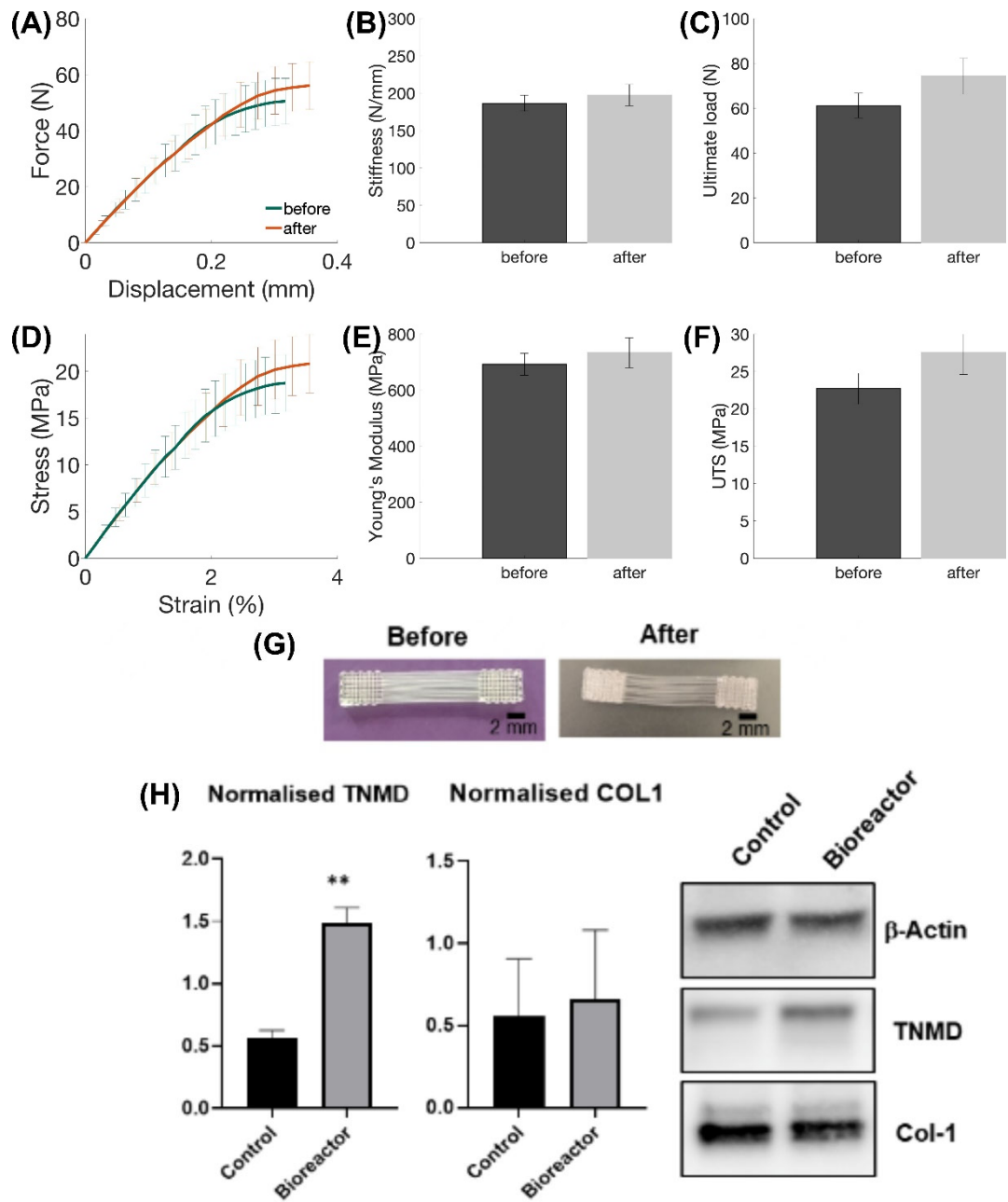

**Figure 8.** Mechanical properties of the scaffolds (ligament part 15mm to fit the bioreactor) before and after 6 days in the bioreactor (A) Force displacement curve, (B) stiffness (C) ultimate load, (D) Stress- strain curve, (E) Young's Modulus, (F) Ultimate Tensile Stress (n=5) and (G) images of the scaffolds before and after bioreactor and (H) protein quantification of TNMD and COL1 Western blot studies of tendon-derived stem cells (β-actin is used as a control) (n=3). Stars denote statistical significance (p<0.05).

## 4. Discussion

This study reported on the additive manufacturing of a bone-ligament-bone PET scaffold for SLIL reconstruction. Several technologies have been developed to address SLIL reconstruction, but only very recently has bone-ligament-bone 3D printed scaffolds been proposed as a potential alternative to the current surgical treatment [17, 21]. However, the use of biodegradable construct materials represents a potential drawback for mid and long-term stabilisation of the wrist as a significant decrease in mechanical properties is usually observed in the early stages of the degradative process. Therefore, a non-resorbable polymer with enhanced mechanical properties was assessed for suitability in SLIL reconstruction in the present study. PET is a biomaterial which has previously been utilised in orthopaedics and especially for anterior cruciate ligament reconstruction with comparable rate of clinical success to standard reconstruction techniques [28] (even if several long-term complications have been reported [40]). In addition, PET is a commercially available filament material for 3D-printing and given its inherent mechanical properties it is an interesting candidate for the reconstruction of SLIL.

Optimisation of printing conditions was performed, showing a strong influence of the printing temperature and speeds on structure and reproducibility of the scaffolds. Optimal printing temperature was 200 °C and faster print speeds allowed better control of strut alignment within the ligament compartment. Thus, 200 °C and 600mm/s were the optimal printing conditions. Tensile mechanical properties were evaluated under quasi-static conditions and showed the PET scaffolds printed at 200C and 600mm/min had mechanical properties suitable for SLIL reconstruction, i.e., tensile stiffness in the range of 250 N/mm and ultimate load of about 110N. The stiffness and ultimate load reported in this present study are significantly higher compared to those previously published for a polycaprolactone bone-ligament-bone scaffold, where stiffness was ~100 N/mm and ultimate load <80 N [17, 21]. Interestingly, mechanical properties of the PET scaffold were in the mid-range of those reported for the native SLIL in the literature (from 30 to 250 N/mm) [41, 42]. In addition, the physiological loading experienced by the SLIL has been estimated to be in the 20-50 N range for normal motion [43, 44], which is well below the yield force of the PET bone-ligament-bone scaffold indicating that physiological deformation would remain in the elastic domain of the scaffold. This suggests the scaffold would be stable for normal carpal physiological movement and may also potentially better withstand higher forces generated by accidental falls on the hands compared to other polymers.

The utilisation of 3D-printed BLB scaffold for SLIL regeneration potentially provides significant advantages over structures such as braided [45] and knitted or woven [46] scaffolds. Indeed, the increased porosity ranging from 55 to 70% (depending on the compartment) enables enhanced and more rapid tissue colonisation and vascularisation especially when compared to dense braided scaffolds[47]. The field of ligament and tendon tissue engineering has also seen the emergence of electrospun scaffolds as potential candidates for tissue regeneration. While the technology can provide tissue guidance using topographical cues in the form of aligned [48] or wavy fibrous patterns [49, 50], their mechanical properties rarely reached those required for the physiological loading of human ligaments.

The bioreactor testing also demonstrated the PET bone-ligament-bone scaffold could withstand cyclic deformation in the physiological range of the SLIL without experiencing irreversible plastic deformation. Although our previous report investigated the fatigue behaviour of the PCL bone-ligament-bone scaffold, this assessment was limited to one thousand cycles and therefore only provided a preliminary validation of the maintenance of mechanical properties across repeated deformation [17]. In the present study, the PET scaffold underwent 0.25 Hz cyclic testing for 8 hours per day of over 6 days, i.e., ~43,000 cycles, and demonstrated excellent recovery from deformation. This was further confirmed by mechanical testing, revealing no significant decreases in stiffness, Young's Modulus, and ultimate load before and after the cyclic bioreactor loading protocol. A limitation of this study is that fatigue of the bone-ligament-bone scaffold was only tested uniaxially, and as such does not capture the complexity of the physiological loading. Nevertheless, it is a step forward demonstrating the potential of this multiphasic bone-ligament-bone construct toward clinical translation.

The outcome of the iFEA modelling focused on fitting material parameters, such as Young's modulus and Poisson's ratio, to accurately describe elastic properties of the PET material. Based on mechanical data obtained from the PET rectangular scaffolds, we may speculate a U-shape scaffold made from PET may have enhanced mechanical properties compared to a scaffold made from PCL as we previously reported [22]. However, further studies that consider the viscoelastic properties of PET must be performed to fully compare both polymers for this application. As this initial work was mostly focussed on the optimisation of parameters, further studies must be completed to model the damage mechanism of the ligament-scaffold. An FEA could be used to predict the mechanical properties of these scaffolds and can be adapted for various polymeric materials and scaffolds morphologies under different physiological loading conditions [51]. Although the PET scaffold may display interesting mechanical and fatigue properties for the stabilisation of the scapholunate joint, a limitation may come from a potential foreign body reaction to the PET materials due to ligament fraying as it has been already observed in several cases for the PET LARS ligament [52, 53] with some cases of failure [54] for however much higher biomechanical stimulation (in range of several hundreds of Newtons) [55]. From a biomechanical perspective, the use of a PET scaffold is still a promising approach for this application.

PET biomaterial is known to be relatively bioinert, therefore a regenerative strategy would benefit of increasing the scaffold bioactivity and biofunctionality. The literature reports several approaches for enhancing the PET biological performance [32, 56] and this is generally attained by performing various surface treatments from calcium phosphate deposition [57, 58], covalent grafting of polystyrene sodium sulphonate [29, 59] dopamine mediated grafting of gelatine [60]. These approaches of surface biofunctionalisation had also the additional effect of increasing tissue integration preventing ligament prothesis fraying and therefore the initiation of a strong foreign body reaction. Our strategy involved the utilisation of plasma treatment, a method commonly used in the biomedical industry, in order to perform surface modification of the scaffold was performed using plasma treatment for altering the surface chemistry of the PET-3D-printed scaffold. This simple treatment increased the amount of C-O and C=O on the surface, which is essential for enhancing cell attachment on the scaffold [61]. Indeed, plasma treatment is known to improve the bioactivity of scaffolds, promoting cell attachment, thus tissue healing [62]. In addition, the treatment resulted in profound changes in the fluid penetration ability within the scaffold as an increased hydrophilicity was observed, which is consistent with the literature [63]. The present study subsequently assessed two different techniques, both using cells supplementation, for enhancing the healing potential of the PET 3D-printed construct. The first approach was the addition of cell spheroids as they have a demonstrated and enhanced differentiation capacity over 2D culture [64, 65]. A technical limitation is nevertheless the seeding of cell spheroid in a highly porous structure with high pore sizes and interconnectivity. This was circumvented by the utilisation of GelMA as a delivery carrier for the spheroids. The plasma treatment ensured the penetration of the GelMA precursor solution through the porous network of the scaffold prior to crosslinking, thus resulting in a homogenous distribution of the cell spheroids across the construct. The approach of combining an hydrogel and polymeric scaffold had been explored to reinforce the properties of the hydrogel [66] and also to enhance the biological performance of the scaffold, as the cells were able to fully colonised the interstitial volume. Interestingly, the cells that migrated out of the spheroid followed the architectural features of the scaffold and adopted preferentially an elongated morphology in the longitudinal direction of the printed PET filament in the ligament compartment, indicating a certain level of architectural guidance of the scaffold as previously reported for 3D-printed bone-ligament-bone scaffolds [17, 21].

The second approach to increase the biofunctionability of the scaffold was to directly seed cells in the scaffold prior to a mechanically induced in vitro maturation phase in a bioreactor. This demonstrated an increase in the protein expression of a well-established tenocyte marker after 6 days, confirming the higher commitment of the cells in fibroblastic differentiation which is consistent with previous reports [38, 39, 67]. Although the use of a bioreactor has been previously proposed for enhancing regenerative outcomes [68, 69], it remains expensive and potentially problematic for clinical translation due to the increased regulatory burden and costs. Therefore, use of spheroids may provide a more cost-effective manner to enhance the regenerative capacity of the PET bone-ligament-bone scaffold compared to bioreactor enhancement.

Similarly to our previous studies for SLIL regeneration, the BLB PET-scaffold requires further in vivo testing using more physiologically relevant animal models such as the medial collateral ligament lapine model we previously developed [21]. This will investigate the bone integration and ligament regeneration in a relevant physiological load bearing situation. Our previous study demonstrated the suitability of this model to investigate the regeneration of the ligament fibre insertion transitional zone in the newly formed bone.

## 5. Conclusions

A bone-ligament-bone scaffold was manufactured from PET for use in SLIL reconstruction. The scaffold had excellent tensile properties and appears capable of withstanding physiological loading experienced by the native SLIL. A proof of concept established several approaches for enhancing the biofunctionality of scaffold, of which the delivery of cell spheroid seemed most promising for the regeneration of the SLIL.

## 6. Acknowledgments

The authors acknowledge the facilities and the scientific and technical assistance of the Australian Microscopy & Microanalysis Research Facility at the Centre for Microscopy and Microanalysis, The University of Queensland. The authors also acknowledge acknowledge the Translational Research Institute (TRI) for providing the excellent research environment and core facilities (Kamil Sokolowski, Preclinical Imaging) that enabled this research. The authors also acknowledge the Lions Club of Australia and the Mater Foundation for funding the Skyscan 1272 Micro-CT. This research was funded by an Australian MTPConnect Biomedical Technology Horizons grant to RB, DGL, CV, DJS and MHZ, and grants in support from Griffith University.

## 7. References

[1] R.A. Berger, The gross and histologic anatomy of the scapholunate interosseous ligament, J. Hand Surg. Am. 21 (1996) 170–178.
[2] L. Reissner, O. Politikou, G. Fischer, M. Calcagni, In-vivo three-dimensional motion analysis of the wrist during dart-throwing motion after midcarpal fusion and radioscapholunate fusion, Journal of Hand Surgery (European Volume) 45(5) (2020) 501-507.
[3] G.H. Brigstocke, A. Hearnden, C. Holt, G. Whatling, In-vivo confirmation of the use of the dart thrower's motion during activities of daily living, J Hand Surg Eur Vol 39(4) (2014) 373-8.
[4] S. Sen, S. Talwalkar, Acute and chronic scapholunate ligament instability, Orthopaedics and Trauma 31(4) (2017) 266-273.
[5] A. Schweizer, R. Steiger, Long-term results after repair and augmentation ligamentoplasty of rotatory subluxation of the scaphoid, J Hand Surg Am 27(4) (2002) 674-84.
[6] C.E. Kuo, S.W. Wolfe, Scapholunate instability: current concepts in diagnosis and management, J Hand Surg Am 33(6) (2008) 998-1013.
[7] A. Izadpanah, S. Kakar, Acute Scapholunate Ligament Injuries: Current Concepts, Operative Techniques in Sports Medicine 24(2) (2016) 108-116.
[8] A. Kitay, S.W. Wolfe, Scapholunate instability: Current concepts in diagnosis and management, J. Hand Surg. Am. 37(10) (2012) 2175–2196.
[9] R. Endress, C.Y.L. Woon, S.J. Farnebo, A. Behn, J. Bronstein, H. Pham, X. Yan, S.S. Gambhir, J. Chang, Tissue-engineered Collateral Ligament Composite Allografts for Scapholunate Ligament Reconstruction: An Experimental Study, J. Hand Surg. Am. 37(8) (2012) 1529-1537.
[10] S.K. Lee, D.A. Zlotolow, A. Sapienza, R. Karia, J. Yao, Biomechanical comparison of 3 methods of scapholunate ligament reconstruction, J Hand Surg Am 39(4) (2014) 643-50.
[11] E. Rohman, J. Agel, M. Putnam, J. Adams, Scapholunate interosseous ligament injuries: A retrospective review of treatment and outcomes in 82 wrists, J Hand Surg Am 39(10) (2014) 2020-2026.
[12] N. Pauchard, A. Dederichs, J. Segret, S. Barbary, F. Dap, G. Dautel, The role of three-ligament tenodesis in the treatment of chronic scapholunate instability, J Hand Surg Eur Vol 38(7) (2013) 758-66.

[13] G.A. Brunelli, G.R. Brunelli, A new technique to correct carpal instability with scaphoid rotary subluxation: a preliminary report, J Hand Surg Am 20(3 Pt 2) (1995) S82-5.
[14] J.K. Andersson, Treatment of scapholunate ligament injury: Current concepts, EFORT Open Rev 2(9) (2017) 382-393.
[15] H. Lui, S. Kakar, Arthroscopic-Assisted Volar Scapholunate Capsulodesis: A New Technique, The Journal of Hand Surgery 47 (2022).
[16] M. Endres, D. Hutmacher, A. Salgado, C. Kaps, J. Ringe, R. Reis, M. Sittinger, A. Brandwood, J.-T. Schantz, Osteogenic induction of human bone marrow-derived mesenchymal progenitor cells in novel synthetic polymer-hydrogel matrices, Tissue engineering 9(4) (2003) 689-702.
[17] H. Lui, R. Bindra, J. Baldwin, S. Ivanovski, C. Vaquette, Additively Manufactured Multiphasic Bone–Ligament–Bone Scaffold for Scapholunate Interosseous Ligament Reconstruction, Advanced Healthcare Materials 8(14) (2019).
[18] H. Lui, C. Vaquette, R. Bindra, Tissue Engineering in Hand Surgery: A Technology Update, J Hand Surg Am 42(9) (2017) 727-735.
[19] X. Jiang, Y. Kong, M. Kuss, J. Weisenburger, H. Haider, R. Harms, W. Shi, B. Liu, W. Xue, J. Dong, J. Xie, P. Streubel, B. Duan, 3D bioprinting of multilayered scaffolds with spatially differentiated ADMSCs for rotator cuff tendon-to-bone interface regeneration, Applied Materials Today 27 (2022) 101510.
[20] S. Chae, Y.-J. Choi, D.-W. Cho, Mechanically and biologically promoted cell-laden constructs generated using tissue-specific bioinks for tendon/ligament tissue engineering applications, Biofabrication 14(2) (2022) 025013.
[21] H. Lui, C. Vaquette, J.M. Denbeigh, R. Bindra, S. Kakar, A.J. van Wijnen, Multiphasic scaffold for scapholunate interosseous ligament reconstruction: A study in the rabbit knee, Journal of Orthopaedic Research in press (2021).
[22] N. Perevoshchikova, K.M. Moerman, B. Akhbari, R. Bindra, J.N. Maharaj, D.G. Lloyd, M. Gomez Cerezo, A. Carr, C. Vaquette, D.J. Saxby, Finite element analysis of the performance of additively manufactured scaffolds for scapholunate ligament reconstruction, PLoS One 16(11) (2021) e0256528.
[23] C.X.F. Lam, S.H.T. Teoh, D.W. Hutmacher, Comparison of the degradation of polycaprolactone and polycaprolactone-(β-tricalcium phosphate) scaffolds in alkaline medium, Polymer International 56 (2007) 718–728.
[24] C.X. Lam, D.W. Hutmacher, J.T. Schantz, M.A. Woodruff, S.H. Teoh, Evaluation of polycaprolactone scaffold degradation for 6 months in vitro and in vivo, J Biomed Mater Res A 90(3) (2009) 906-19.
[25] X.H. Zong, Z.G. Wang, B.S. Hsiao, B. Chu, J.J. Zhou, D.D. Jamiolkowski, E. Muse, E. Dormier, Structure and morphology changes in absorbable poly(glycolide) and poly(glycolide-co-lactide) during in vitro degradation, Macromolecules 32(24) (1999) 8107-8114.
[26] C. Vaquette, S. Slimani, C.J.F. Kahn, N. Tran, R. Rahouadj, X. Wang, A poly(lactic-co-glycolic acid) knitted scaffold for tendon tissue engineering: An in vitro and in vivo study, Journal of Biomaterials Science, Polymer Edition 21(13) (2010) 1737-1760.
[27] S. Li, H. Garreau, M. Vert, Structure-property relationships in the case of the degradation of massive poly(α-hydroxy acids) in aqueous media - Part 3 Influence of the morphology of poly(l-lactic acid), Journal of Materials Science: Materials in Medicine 1(4) (1990) 198-206.

[28] Z. Machotka, I. Scarborough, W. Duncan, S. Kumar, L. Perraton, Anterior cruciate ligament repair with LARS (ligament advanced reinforcement system): a systematic review, Sports Med Arthrosc Rehabil Ther Technol 2 (2010) 29.
[29] C. Vaquette, V. Viateau, S. Guerard, F. Anagnostou, M. Manassero, D.G. Castner, V. Migonney, The effect of polystyrene sodium sulfonate grafting on polyethylene terephthalate artificial ligaments on in vitro mineralisation and in vivo bone tissue integration, Biomaterials 34(29) (2013) 7048-7063.
[30] K. Eng, M. Wagels, S.K. Tham, Cadaveric scapholunate reconstruction using the ligament augmentation and reconstruction system, Journal of wrist surgery 3(3) (2014) 192-197.
[31] H. Li, S. Chen, Biomedical coatings on polyethylene terephthalate artificial ligaments, J Biomed Mater Res A 103(2) (2015) 839-45.
[32] J. Cai, L. Zhang, J. Chen, S. Chen, Silk fibroin coating through EDC/NHS crosslink is an effective method to promote graft remodeling of a polyethylene terephthalate artificial ligament, Journal of Biomaterials Applications 33(10) (2019) 1407-1414.
[33] K. Moerman, GIBBON: The Geometry and Image-Based Bioengineering add-On, Journal of Open Source Software 3 (2018) 506.
[34] S.A. Maas, B.J. Ellis, G.A. Ateshian, J.A. Weiss, FEBio: finite elements for biomechanics, J Biomech Eng 134(1) (2012) 011005.
[35] K. Levenberg, A method for the solution of certain non-linear problems in least squares, Quarterly of Applied Mathematics 2(2) (1944) 164-168.
[36] P.A. Tran, H.T. Nguyen, P.J. Hubbard, H.P. Dang, D.W. Hutmacher, Mineralization of plasma treated polymer surfaces from super-saturated simulated body fluids, Materials Letters 230 (2018) 12-15.
[37] H. Poli, A.L. Mutch, A. A, S. Ivanovski, C. Vaquette, D.G. Castner, M.N. Gómez-Cerezo, L. Grøndahl, Evaluation of surface layer stability of surface-modified polyester biomaterials, Biointerphases 15(6) (2020) 061010.
[38] Z. Chen, P. Chen, R. Ruan, L. Chen, J. Yuan, D. Wood, T. Wang, M.H. Zheng, Applying a Three-dimensional Uniaxial Mechanical Stimulation Bioreactor System to Induce Tenogenic Differentiation of Tendon-Derived Stem Cells, J Vis Exp (162) (2020).
[39] Z. Chen, P. Chen, R. Ruan, M. Zheng, In Vitro 3D Mechanical Stimulation to Tendon-Derived Stem Cells by Bioreactor, Methods Mol Biol 2436 (2022) 135-144.
[40] T.M. Tiefenboeck, E. Thurmaier, M.M. Tiefenboeck, R.C. Ostermann, J. Joestl, M. Winnisch, M. Schurz, S. Hajdu, M. Hofbauer, Clinical and functional outcome after anterior cruciate ligament reconstruction using the LARS™ system at a minimum follow-up of 10years, The Knee 22(6) (2015) 565-568.
[41] E.Q. Pang, N. Douglass, A. Behn, M. Winterton, M.J. Rainbow, R.N. Kamal, Tensile and Torsional Structural Properties of the Native Scapholunate Ligament, J Hand Surg Am 43(9) (2018) 864.e1-864.e7.
[42] P. Rajan, C. Day, Scapholunate interosseous ligament anatomy and biomechanics, J Hand Surg Am 40(8) (2015) 1692-1702.
[43] C. Dimitris, F.W. Werner, D.A. Joyce, B.J. Harley, Force in the Scapholunate Interosseous Ligament During Active Wrist Motion, J Hand Surg Am 40(8) (2015) 1525-33.
[44] L. Scordino, F.W. Werner, B.J. Harley, Force in the Scapholunate Interosseous Ligament During 2 Simulated Pushup Positions, J Hand Surg Am 41(5) (2016) 624-9.

[45] H.H. Lu, J.A. Cooper, S. Manuel, J.W. Freeman, M.A. Attawia, F.K. Ko, C.T. Laurencin, Anterior cruciate ligament regeneration using braided biodegradable scaffolds: in vitro optimization studies, Biomaterials 26(23) (2005) 4805-4816.
[46] S. Wu, Y. Wang, P.N. Streubel, B. Duan, Living nanofiber yarn-based woven biotextiles for tendon tissue engineering using cell tri-culture and mechanical stimulation, Acta Biomater 62 (2017) 102-115.
[47] J.A. Cooper, J.S. Sahota, W.J. Gorum, J. Carter, S.B. Doty, C.T. Laurencin, Biomimetic tissue-engineered anterior cruciate ligament replacement, Proceedings of the National Academy of Sciences 104(9) (2007) 3049-3054.
[48] C. Vaquette, C. Kahn, C. Frochot, C. Nouvel, J.L. Six, N. De Isla, L.H. Luo, J. Cooper-White, R. Rahouadj, X.O. Wang, Aligned poly(L-lactic-co-e-caprolactone) electrospun microfibers and knitted structure: A novel composite scaffold for ligament tissue engineering, Journal of Biomedical Materials Research Part A 94A(4) (2010) 1270-1282.
[49] S. Wu, J. Liu, Y. Qi, J. Cai, J. Zhao, B. Duan, S. Chen, Tendon-bioinspired wavy nanofibrous scaffolds provide tunable anisotropy and promote tenogenesis for tendon tissue engineering, Materials Science and Engineering: C 126 (2021) 112181.
[50] M. Gwiazda, S. Kumar, W. Świeszkowski, S. Ivanovski, C. Vaquette, The effect of melt electrospun writing fiber orientation onto cellular organization and mechanical properties for application in Anterior Cruciate Ligament tissue engineering, Journal of the Mechanical Behavior of Biomedical Materials 104 (2020).
[51] A.A. Soufivand, N. Abolfathi, S.A. Hashemi, S.J. Lee, Prediction of mechanical behavior of 3D bioprinted tissue-engineered scaffolds using finite element method (FEM) analysis, Additive Manufacturing 33 (2020) 101181.
[52] Y. Du, H. Dai, Z. Wang, D. Wu, C. Shi, T. Xiao, Z. Li, A case report of traumatic osteoarthritis associated with LARS artificial ligament use in anterior cruciate ligament reconstruction, BMC Musculoskeletal Disorders 21(1) (2020) 745.
[53] Z.P. Sinagra, A. Kop, M. Pabbruwe, J. Parry, G. Clark, Foreign Body Reaction Associated With Artificial LARS Ligaments: A Retrieval Study, Orthop J Sports Med 6(12) (2018) 2325967118811604.
[54] H. Li, Z. Yao, J. Jiang, Y. Hua, J. Chen, Y. Li, K. Gao, S. Chen, Biologic failure of a ligament advanced reinforcement system artificial ligament in anterior cruciate ligament reconstruction: a report of serious knee synovitis, Arthroscopy 28(4) (2012) 583-6.
[55] K.L. Markolf, S.R. Jackson, B. Foster, D.R. McAllister, ACL forces and knee kinematics produced by axial tibial compression during a passive flexion–extension cycle, Journal of Orthopaedic Research 32(1) (2014) 89-95.
[56] Y.-L. Kuo, F.-C. Kung, C.-L. Ko, A. Okino, T.-C. Chiang, J.-Y. Guo, S.-Y. Chen, Tailoring surface properties of polyethylene terephthalate by atmospheric pressure plasma jet for grafting biomaterials, Thin Solid Films 709 (2020) 138152.
[57] J. Cai, Q. Zhang, J. Chen, J. Jiang, X. Mo, C. He, J. Zhao, Electrodeposition of calcium phosphate onto polyethylene terephthalate artificial ligament enhances graft-bone integration after anterior cruciate ligament reconstruction, Bioactive Materials 6(3) (2021) 783-793.
[58] C.C. Tai, C.C. Huang, B.H. Chou, C.Y. Chen, S.Y. Chen, Y.H. Huang, J.S. Sun, Y.H. Chao, Profiled polyethylene terephthalate filaments that incorporate collagen and calcium phosphate enhance ligamentisation and bone formation, Eur Cell Mater 43 (2022) 252-266.
[59] M. Ciobanu, A. Siove, V. Gueguen, L.J. Gamble, D.G. Castner, V. Migonney, Radical graft polymerization of styrene sulfonate on poly(ethylene terephthalate) films

for ACL applications: "grafting from" and chemical characterization, Biomacromolecules 7 (2006) 755-760.
[60] E.D. Giol, D. Schaubroeck, K. Kersemans, F. De Vos, S. Van Vlierberghe, P. Dubruel, Bio-inspired surface modification of PET for cardiovascular applications: Case study of gelatin, Colloids and Surfaces B: Biointerfaces 134 (2015) 113-121.
[61] I. Junkar, A. Vesel, U. Cvelbar, M. Mozetič, S. Strnad, Influence of oxygen and nitrogen plasma treatment on polyethylene terephthalate (PET) polymers, Vacuum 84(1) (2009) 83-85.
[62] S. Sreeja, C.V. Muraleedharan, P.R.H. Varma, G.S. Sailaja, Surface-transformed osteoinductive polyethylene terephthalate scaffold as a dual system for bone tissue regeneration with localized antibiotic delivery, Mater Sci Eng C Mater Biol Appl 109 (2020) 110491.
[63] S.B. Amor, M. Jacquet, P. Fioux, M. Nardin, XPS characterisation of plasma treated and zinc oxide coated PET, Applied Surface Science 255(9) (2009) 5052-5061.
[64] Y. Yamaguchi, J. Ohno, A. Sato, H. Kido, T. Fukushima, Mesenchymal stem cell spheroids exhibit enhanced in-vitro and in-vivo osteoregenerative potential, BMC Biotechnology 14(1) (2014) 105.
[65] L. Li, X. Liu, B. Gaihre, Y. Li, L. Lu, Mesenchymal stem cell spheroids incorporated with collagen and black phosphorus promote osteogenesis of biodegradable hydrogels, Materials Science and Engineering: C 121 (2021) 111812.
[66] J. Visser, F.P. Melchels, J.E. Jeon, E.M. Van Bussel, L.S. Kimpton, H.M. Byrne, W.J. Dhert, P.D. Dalton, D.W. Hutmacher, J. Malda, Reinforcement of hydrogels using three-dimensionally printed microfibres, Nature Communications 6 (2015) 6933.
[67] M. Lee, B.M. Wu, 17 - Tissue engineering for ligament and tendon repair, in: C. Archer, J. Ralphs (Eds.), Regenerative Medicine and Biomaterials for the Repair of Connective Tissues, Woodhead Publishing2010, pp. 419-435.
[68] C.J.F. Kahn, C. Vaquette, R. Rahouadj, X. Wang, A novel bioreactor for ligament tissue engineering, Bio-Medical Materials and Engineering 18(4-5) (2008) 283-287.
[69] C.P. Laurent, C. Vaquette, C. Martin, E. Guedon, X. Wu, A. Delconte, D. Dumas, S. Hupont, N. De Isla, R. Rahouadj, X. Wang, Towards a tissue-engineered ligament: Design and preliminary evaluation of a dedicated multi-chamber tension-torsion bioreactor, Processes 2(1) (2014) 167-179.